\documentclass[9pt,twocolumn,twoside]{pnas-new}
\usepackage{enumitem}
\setitemize[0]{itemsep=0pt,partopsep=0pt,parsep=0pt,topsep=0pt}

\templatetype{pnasresearcharticle} 
\usepackage{pdfpages}

\title{Quantum critical scaling and fluctuations in Kondo lattice materials}

\author[a,b,c,1]{Yi-feng Yang}
\author[d,2]{David Pines} 
\author[e]{Gilbert Lonzarich}

\affil[a]{Beijing National Laboratory for Condensed Matter Physics and Institute of Physics, Chinese Academy of Sciences, Beijing 100190, China}
\affil[b]{Collaborative Innovation Center of Quantum Matter, Beijing 100190, China}
\affil[c]{School of Physical Sciences, University of Chinese Academy of Sciences, Beijing 100190, China}
\affil[d]{Santa Fe Institute, Santa Fe, NM 87501, USA}
\affil[e]{Cavendish Laboratory, Department of Physics, Cambridge University, Cambridge CB3 0HE, UK}
\leadauthor{Yang} 

\significancestatement{
Quantum critical materials contain quantum critical points (QCP) associated with continuous phase transitions between ordered states at zero temperature that give rise to strong quantum fluctuations of order parameter fields. Their coupling to conduction electrons in metals typically yields logarithmic or power law scaling in their electronic properties. Here we study a particularly interesting class of quantum critical materials, Kondo lattice materials, composed of interacting local moments and conduction electrons. We propose a phenomenological framework based on the interplay of the fluctuations associated with their antiferromagnetic and delocalization QCPs and show that it explains the many anomalous scaling properties measured in these materials.}

\authorcontributions{All authors contribute to the idea and writing of the manuscript.}
\authordeclaration{The authors declare no conflict of interest.}
\correspondingauthor{\textsuperscript{1,2}To whom correspondence should be addressed. E-mail: yifeng@iphy.ac.cn; david.pines@gmail.com}


\begin{abstract}
We propose a new phenomenological framework for three classes of Kondo lattice materials that incorporates the interplay between the fluctuations associated with the antiferromagnetic quantum critical point and those produced by the hybridization quantum critical point that marks the end of local moment behavior. We show that these fluctuations give rise to two distinct regions of quantum critical scaling: hybridization fluctuations are responsible for the logarithmic scaling in the density of states of the heavy electron Kondo liquid that emerges below the coherence temperature $T^*$; while the unconventional power law scaling in the resistivity that emerges at lower temperatures below $T_{QC}$ may reflect the combined effects of hybridization and antiferromagnetic quantum critical fluctuations. Our framework is supported by experimental measurements on CeCoIn$_5$, CeRhIn$_5$ and other heavy electron materials.
\end{abstract}

\dates{This manuscript was compiled on \today}

\begin{document}

\verticaladjustment{-2pt}

\maketitle
\thispagestyle{firststyle}
\ifthenelse{\boolean{shortarticle}}{\ifthenelse{\boolean{singlecolumn}}{\abscontentformatted}{\abscontent}}{}

\dropcap{H}eavy electron materials stand out in the correlated electron family because of the extraordinary variety of quantum mysteries these present. In addition to exhibiting two ordered states at low temperatures, antiferromagnetism and superconductivity, that can coexist, essentially nothing about their higher temperature normal state behavior is what one finds in "normal" materials. Not only does the interaction between a lattice of localized f electron magnetic moments and background conduction electrons give rise to the emergence, at a temperature $T^*$ (often called the coherence temperature), of heavy electrons with masses that can become comparable to that of a muon, every other aspect of their normal state behavior produced by that interaction is anomalous.

\begin{figure}[t]
\centerline{{\includegraphics[width=.45\textwidth]{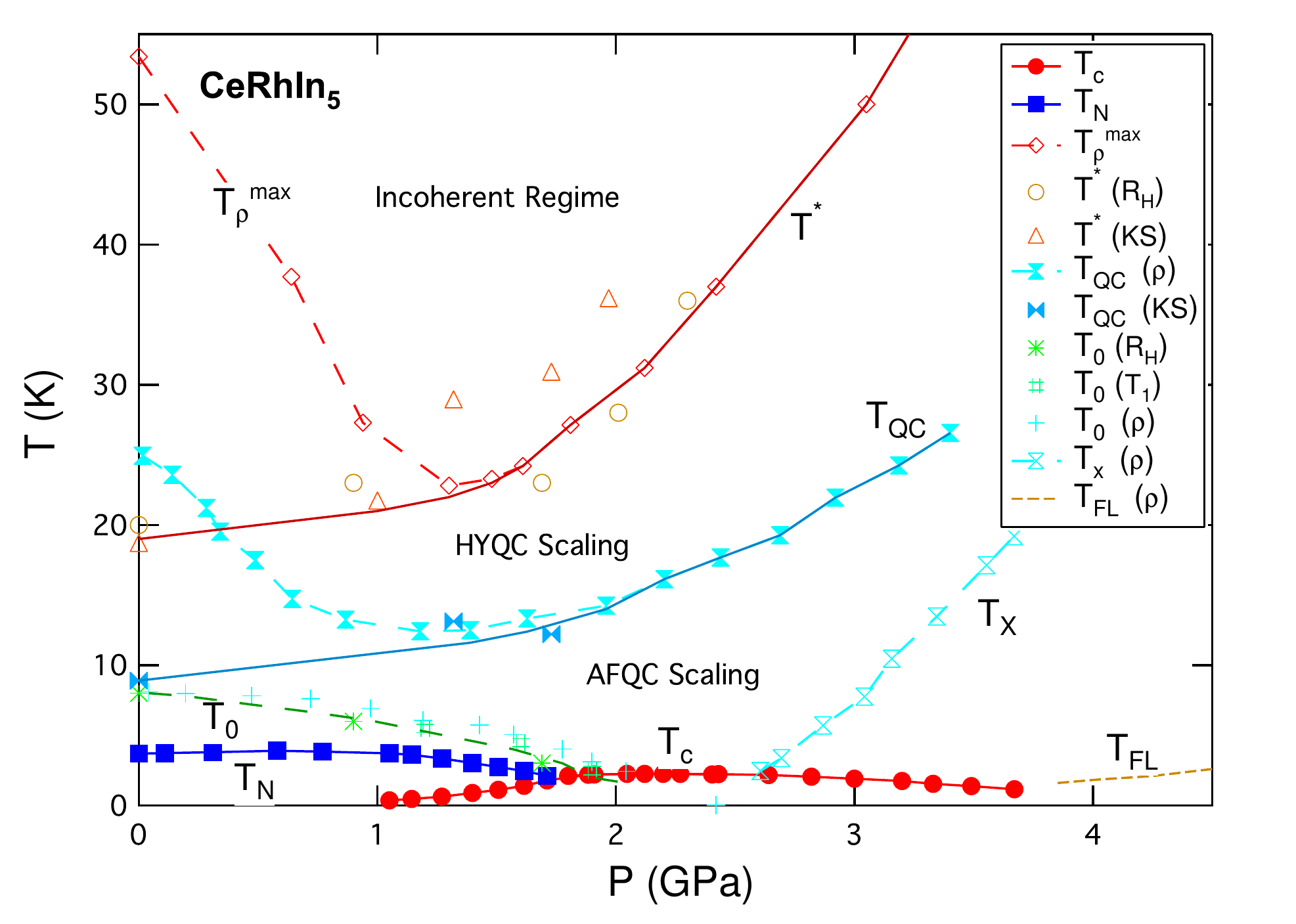}}}
\caption{
{The temperature-pressure phase diagram of CeRhIn$_5$. $T_\rho^{max}$ marks the temperature of the resistivity peak \cite{Park2011}; $T^*$ is determined from the resistivity peak (above 1.5 GPa) and the onset of the Hall and Knight shift anomalies \cite{Yang2012}; $T_{QC}$ is the upper boundary of the power law scaling in the resistivity (above 1.5 GPa) \cite{Park2011} and the breakdown of the Knight shift scaling \cite{Lin2015}; $T_0$ is the lower boundary of the power law scaling in the resistivity (below 2 GPa) \cite{Park2011}, the peak in the Hall coefficient, a pseudo-gap like feature in the spin-lattice relaxation rate, and, at ambient pressure, the onset of antiferromagnetic spin fluctuations seen in neutron scattering experiments \cite{Yang2012}; $T_X$ marks the lower boundary of the power law scaling in the resistivity (above 2.5 GPa); $T_{FL}$ is determined from the Fermi liquid behavior in the resistivity; $T_N$ and $T_c$ are the antiferromagnetic and superconducting transition temperatures, respectively \cite{Park2011}.}
\label{fig1}}
\end{figure}

Experiments on the best studied heavy electron material, CeRhIn$_5$, show that as the temperature and pressure are varied, some five different temperature scales, all well below the crystal field energy levels, are needed to characterize the normal state anomalies depicted in Fig.~\ref{fig1} \cite{Yang2016,Yang2012,Park2011,Lin2015}:
\begin{itemize}[leftmargin=*]
\item{a nuclear magnetic Knight shift that does not follow the measured spin susceptibility below $T^*$.}

\item{a lower limit, $T_{QC}$, on the $\ln T$ universal behavior of the heavy electron density of states that begins at $T^*$.}

\item{a maximum in the magnetic resistivity at $T^{\text{max}}_{\rho}$.}

\item{a lower limit, $T_0$ or $T_X$ depending on the pressure range in which it is studied, on the power law scaling behavior in the resistivity that begins at a temperature, $T_{QC}$.}

\end{itemize}
It is widely believed that the source of these anomalies, and similar ones found in other heavy electron materials, are fluctuations associated with quantum critical points that mark transition between distinct phases of matter at $T=0$. Although there exist microscopic theories of aspects of this quantum critical behavior  [the Hertz-Millis-Moriya model for the spin fluctuation spectrum near an antiferromagnetic quantum critical point (AF QCP) \cite{Hertz1976,Millis1993,Moriya1995}, the Abrahams-W\"olfle model of critical quasiparticles at very low temperatures for materials that are very near an AF QCP \cite{Abrahams2012}; the work of Coleman, Pepin, Senthil, and Si {\sl et al.} on new critical excitations beyond the basic order parameter fluctuations \cite{Coleman2001,Si2001,Senthil2004,Paul2007}; and that of Lonzarich {\sl et al.} suggesting a path forward for an improved microscopic approach to understanding the emergence of heavy electrons in Kondo lattice materials \cite{Lonzarich2017}], these do not explain all the above anomalies, not least because there is at present no microscopic theory of the behavior of the three components (light conduction electrons, heavy electrons, and local moments that have partially hybridized), that exist over much of the phase diagram.
 
We therefore turn to phenomenology in our search for an understanding of the wide range of anomalous behavior and the temperature scales over which it is found. In what follows we show that a careful analysis of the phase diagrams expected from the presence of two competing quantum critical points, one associated with the end of antiferromagnetism, the other with the hybridization-induced end of local moment behavior, together with the phenomenological two-fluid model of that hybridization [for a review see Ref.~\cite{Yang2016}], provides a framework that enables us to arrive at a more complete physical understanding of the anomalous normal state behavior of CeRhIn$_5$, and a number of other  well-studied heavy electron materials.

\section{QCPs and their expected scaling signatures}
In the CeMIn$_5$ (M=Co, Rh, Ir) and similar Kondo lattice materials, we have argued that "collective" hybridization of the local moments against the background conduction electrons begins at the coherence temperature, $T^*$, and is complete along a line of temperatures, $T_L$ \cite{Yang2016}: above $T_L$ we expect to find both local moments whose strength has been reduced by hybridization that can order antiferromagnetically and itinerant heavy electrons that can become superconducting; well below it we will find only heavy electrons that may superconduct. Absent superconductivity, we would then expect to find two distinct quantum critical points: an AF QCP that marks the $T=0$ end of local moment antiferromagnetic behavior and a hybridization quantum critical point (HY QCP) that marks the $T=0$ completion of collective hybridization of local moments. These QCPs can produce quantum critical local moment spin fluctuations and quantum critical heavy electron spin or charge fluctuations, and a key question is the extent to which these QCPs and the scaling behaviors to which they give rise, can be distinguished and identified experimentally.

\begin{figure}[t]
\centerline{{\includegraphics[width=.4\textwidth]{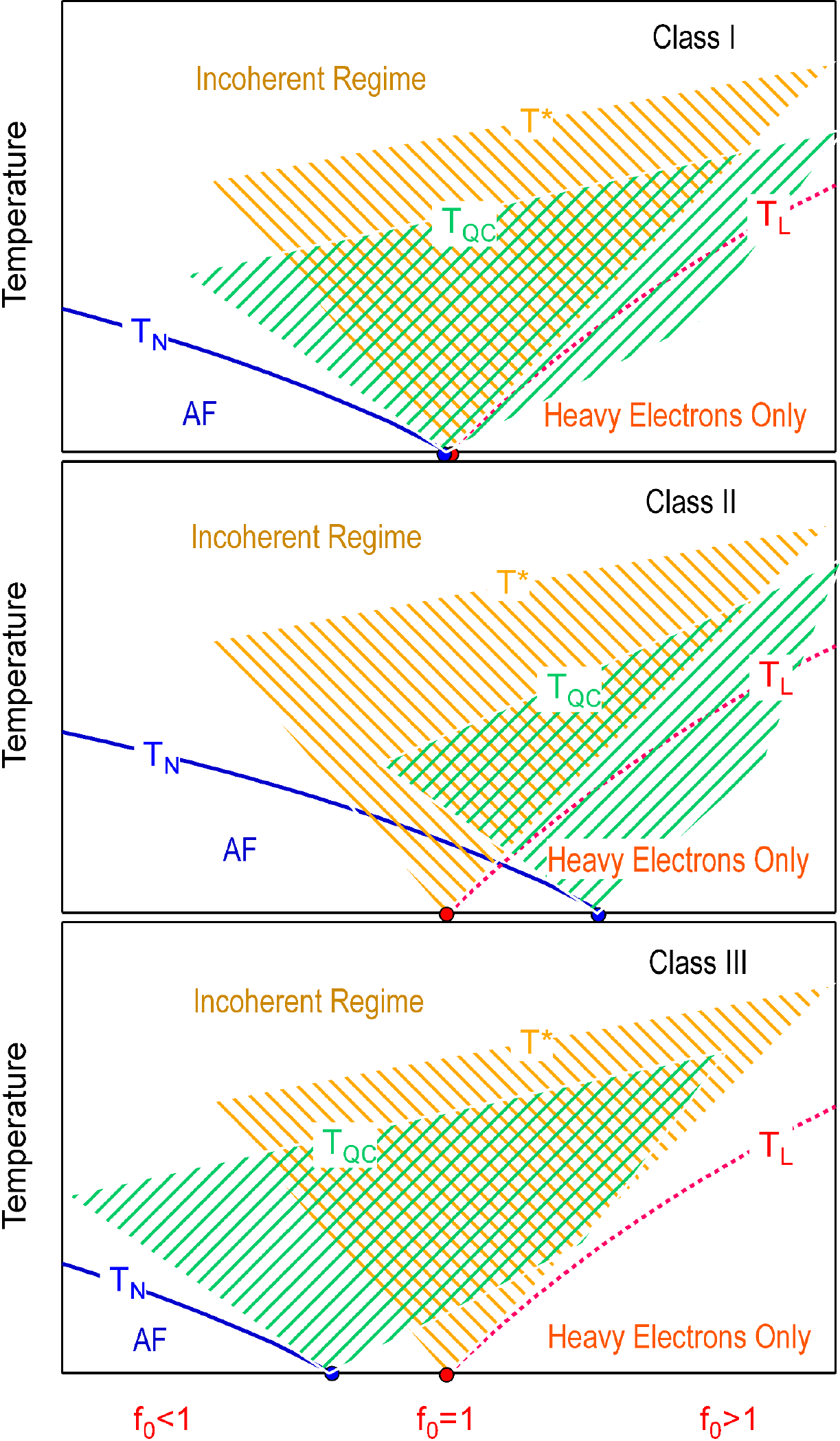}}}
\caption{
{An illustration of the interplay between the quantum critical fluctuations associated with the AF QCP and the HY QCP for three classes of heavy electron materials: in Class I, two QCPs coincide; in Class II, the AF QCP lies inside the heavy electron only phase and the HY QCP is inside the N\'eel phase; in Class III, these separate and one expects a non-magnetic non-Fermi liquid phase in between the two QCPs. The shaded areas illustrate possible ranges of their associated quantum critical fluctuations. $f_0$ is the intrinsic hybridization strength defined in Eq.~\ref{eq2}.}
\label{fig2}}
\end{figure}

Quite generally we may expect to find the three classes of heavy electron antiferromagnetic materials shown in Fig.~\ref{fig2} \cite{Si2010,Coleman2010}. Class I materials are those in which the AF and HY QCPs appear to be identical within experimental error. Class II are those in which the HY QCP lies well within the antiferromagnetic phase; in this case, the AF QCP becomes an itinerant AF QCP associated with the magnetic instability of the itinerant heavy electrons. Class III are those in which that HY QCP lies well outside the antiferromagnetic phase; between the two QCPs there will then be a region in which a nonmagnetic non-Landau heavy electron liquid coexists with incoherent local moments.

We follow Lonzarich {\sl et al.} \cite{Lonzarich2017} in making the assumption that in all three classes we are dealing with two order parameters (and their associated quantum critical points and fluctuations): an HY order parameter describing the build up of the emergent heavy electrons and the more familiar AF order parameter describing the build up of local moment AF order. The magnetic and hybridization quantum critical fluctuations will often not behave independently. For example, we shall see that hybridization can be suppressed and reversed at low temperatures by local moment AF order, causing relocalization of heavy electrons and a corresponding decrease in the hybridization order parameter \cite{Shirer2012}. This relocalization takes place at a temperature slightly above the N\'eel temperature, possibly associated with thermal fluctuations of the AF order parameter. Still another possibility is that the coupling between the magnetic and hybridization quantum critical fluctuations gives rise to unconventional quantum critical scaling in their overlap regime at finite temperatures \cite{Coleman2005,Lohneysen2007,Gegenwart2008}. Moreover, the magnetic quantum critical fluctuations may also penetrate into the region where all f electrons become itinerant, causing a change in the characteristics of the quantum critical scaling from an unconventional type to an itinerant spin density wave (SDW) type. The latter is clearly seen in Class II materials, but may well exist in Class I materials, while in Class III materials, one may expect changes of the scaling exponent when approaching the two separated QCPs with lowering temperature.

In making the plots of the impact of the HY quantum critical fluctuations shown in Fig.~\ref{fig2}, we are making a key assumption about the origins of two parts of the scaling behavior seen in heavy electron liquid that have led it to be called a Kondo liquid (KL): universal scaling behavior, characterized by the energy scale, $T^*$, of the effective order parameter $f(T)$ that measures its strength (see Eq.~\ref{eq2}); and the scaling with $\ln T$ of the intrinsic KL state density seen in uniform magnetic susceptibility and specific heat experiments \cite{Nakatsuji2004,Yang2008}. A central thesis of the present paper is that these two parts represent distinct scaling phenomena of distinguishable physical origins. 

As first shown by Yang {\sl et al.} \cite{Yang2008nature}, $T^*$ is determined by the nearest-neighbor coupling between local moments in the Kondo lattice. Their interaction produces collective hybridization below $T^*$ that is quite different from the single-ion Kondo hybridization (screening) found for isolated magnetic moments. In the present paper we argue that the $\ln T$ scaling behavior seen in the KL state density is brought about by the HY QCP fluctuations (and/or their associated gauge fluctuations) whose influence is cut off above $T^*$.
 
Our main focus in this paper will be on Class I materials; materials belonging to the other two classes are discussed only briefly. It is in fact possible that in Class I materials the localization and magnetic QCPs are never exactly identical, since the combined effects of the HY and AF quantum critical fluctuations may act to move the AF and HY QCPs to opposite directions, reflecting the way in which hybridization fluctuations interfere with long range magnetic order and spin fluctuations interfere with collective hybridization in the vicinity of the putative identical QCP.

\begin{figure}[t]
\centerline{{\includegraphics[width=.45\textwidth]{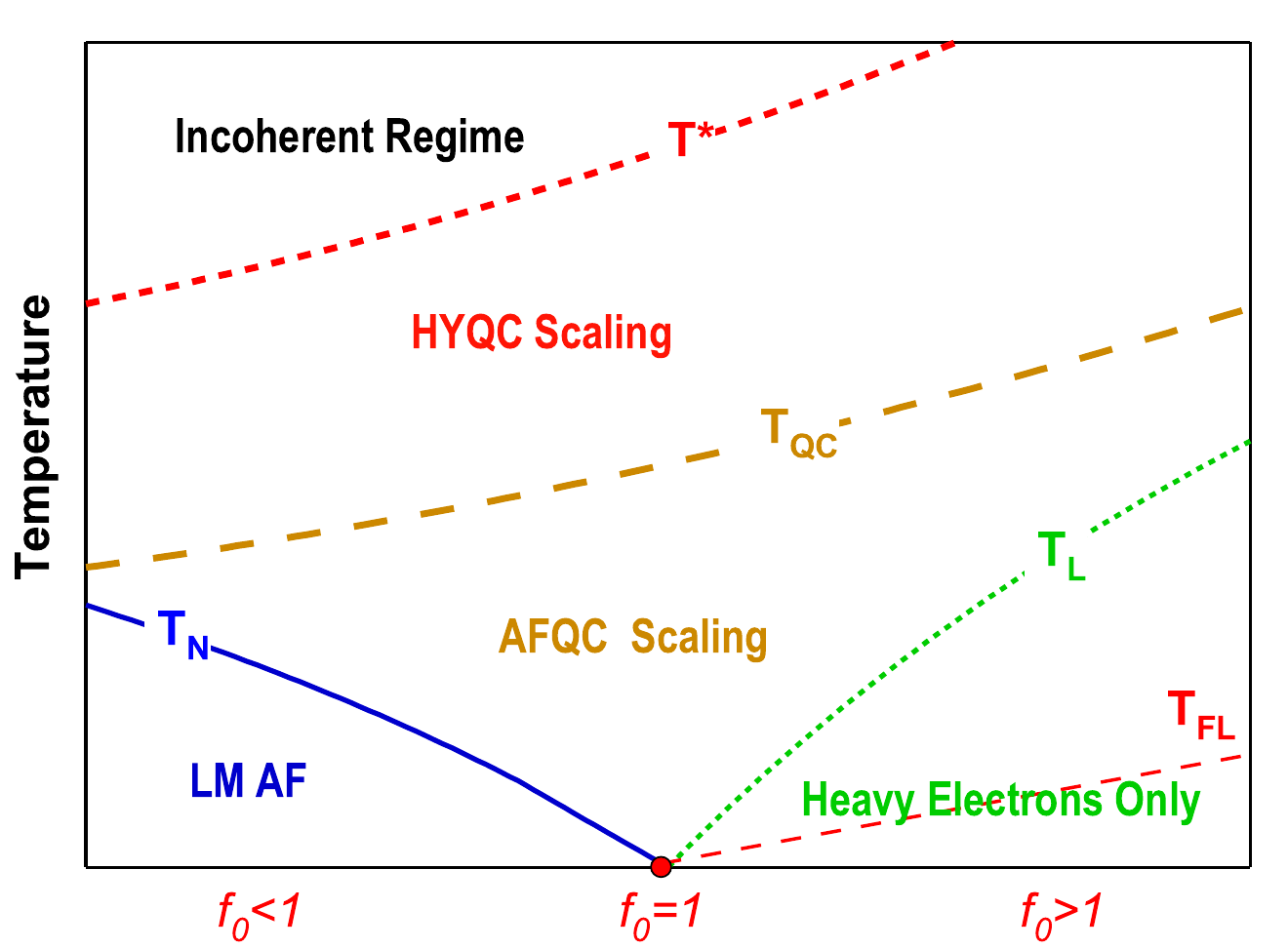}}}
\caption{
{A suggested phase diagram for Class I materials. $T^*$ is the coherence temperature that marks the emergence of the heavy electron liquid whose intrinsic density of states (as defined via the linear coefficient of the specific heat) displays logarithmic scaling behavior brought about by the HY QCP; $T_{QC}$ denotes the boundary between this and the AF quantum critical scaling regime; $T_N$  is the temperature at which the hybridized local moments begin to order; $T_L$ is the hybridization line well below which all f electrons become itinerant; $T_{FL}$ marks the onset of Landau Fermi liquid behavior for the heavy electron liquid.}
\label{fig3}}
\end{figure}

Absent superconductivity, an analysis of a number of experiments on heavy electron materials at comparatively high temperatures ($> 2\ $K) yields the general phase diagram shown in Fig.~\ref{fig3}, in which heavy electrons begin to emerge at a temperature of the order of  $T^*$ as a result of collective hybridization of local moments with the background (light) conduction electrons, and behave like a new quantum state of matter, that exhibits HY quantum critical scaling between $T^*$ and $T_{QC}$. Below $T_{QC}$, although one continues to have coexisting local moments and heavy electrons over much of the phase diagram, the heavy electrons no longer exhibit their KL scaling behavior but potentially display a more dramatic power law divergence because of the proximity of the AF QCP.

Three other important temperature scales are shown there \cite{Yang2012}: $T_N$, the N\'eel temperature at which hybridized local moments begin to exhibit long range magnetic order; $T_L$, the temperature at which collective hybridization of the local moments is nearly complete, so that well below it one finds only heavy electrons; and $T_{FL}$, the temperature at which those heavy electrons begin to exhibit Fermi liquid behavior. 

The phenomenological two-fluid model of the behavior of the coexisting KL and hybridized local moments helps one determine their relative importance for physical phenomena at any pressure or temperature in the phase diagram. For example, the spin susceptibility takes the form 
\begin{equation}
\chi=\left[1-f(T)\right]\chi_{SL}+f(T)\chi_{KL},
\end{equation}
where $\chi_{SL}$ and $\chi_{KL}$ are the intrinsic susceptibility of the spin liquid (hybridized local moments) and the KL, respectively, and $f(T)$, the strength of the KL component, takes the form
\begin{equation}
f(T)=f_0\left(1-T/T^*\right)^{3/2},
\label{eq2}
\end{equation}
where $f_0$, the temperature independent intrinsic "hybridization strength", is the pressure dependent control parameter depicted in Figs.~\ref{fig2} and \ref{fig3}. 

We see that for weakly hybridizing materials, characterized by $f_0<1$, heavy electrons coexist with hybridized local moments until one reaches $T=0$, with the latter ordering antiferromagnetically at $T_N$.  $f_0$ must be unity at the HY QCP at which collective hybridization is complete. For strongly hybridizing ($f_0>1$) materials, that coexistence ends along a line of temperatures, $T_L$, at which the hybridization of local moments is essentially complete. Eq.~\ref{eq2} yields the simple expression, 
\begin{equation}
T_L=T^*\left(1-f_0^{-2/3}\right).
\label{eq3}
\end{equation}
Below $T_L$, these heavy electrons form a quantum liquid that exhibits anomalous quantum critical behavior between $T_L$ and $T_{FL}$, and Landau Fermi liquid behavior below it.

Some additional comments are in order:
\begin{itemize}[leftmargin=*]
\item{Not shown in Fig.~\ref{fig3} is the possible emergence at very low temperatures of a second regime of quantum critical behavior, for which the microscopic theory developed by Abrahams and W\"olfle \cite{Abrahams2012} may be valid.}

\item{Around (and slightly above) the HY QCP there will be a region in which some local moments may be present, but these can reasonably be assumed not to influence the quantum critical behavior of the vast majority of heavy electrons (as required by the Abrahams-W\"olfle model); we arbitrarily take this upper limit to be $\sim$5\%, in which case the two-fluid model tells us that this region will begin at $\sim T^*/30$.}

\item{at $T\sim0$ we expect that well to the left of the HY QCP the Fermi surface will be "small" as it consists of those parts of light electron Fermi surface that have not hybridized with the local moments. To the right of this QCP, the Fermi surface should be "large" as local moments are no longer present. We note that it could be possible that one may observe {\sl effectively} a large Fermi surface in some regions to the left of the HY QCP in which the local moment fraction is too small to preserve the small Fermi surface.}

\item{It is likely that many, if not all, of the lines shown in Fig.~\ref{fig3} do not represent a phase transition, but are indicative of crossover behavior.}

\item{To the extent that one is far from ferromagnetic order, one can neglect the influence of vertex corrections on the static spin susceptibility. Under these circumstances, in both the Kondo liquid and magnetic quantum critical regimes, the uniform magnetic susceptibility $\chi$ and the specific heat $C$ depend only on the heavy electron density of states, $N(0)$, and one expects a temperature independent Wilson ratio.}

\item{For nearly all heavy electron materials existing experiments have yet to provide us with an unambiguous signature for $T_L$, the  temperature below which the Knight shift once again follows the spin susceptibility, since both now originate only in heavy electrons.  Instead one has to rely upon suggestive experimental results such as the crossover in resistivity exponent in CeRhIn$_5$, a maximum in the magneto-resistance in CeCoIn$_5$, a change in the Hall coefficient in YbRh$_2$Si$_2$, and the phenomenological two-fluid expression that relates $T_L$ to the intrinsic hybridization strength, $f_0$, Eq.~\ref{eq3}  \cite{Yang2016,Yang2014}.}

\end{itemize}

\section{CeRhIn$_5$}
CeRhIn$_5$ provides an excellent test of our proposed new framework. At zero magnetic field, the QCPs are hidden by superconductivity \cite{Park2011}. A de Haas-van Alphen (dHvA) experiment in which a strong magnetic field acts to suppress the superconductivity reveals a jump in the Fermi surface upon crossing 2.4 GPa in the high field state \cite{Shishido2005}. This establishes the location of its HY QCP. Since this location is close to the AF QCP obtained by extrapolating the pressure dependence of the N\'eel temperature, CeRhIn$_5$ is likely a Class I material. We now discuss in more detail the phase diagram shown in Fig.~\ref{fig1}:
\begin{itemize}[leftmargin=*]
\setlength{\itemsep}{0pt}
\item{Above $\sim1.5\ $GPa, the values of $T^*$ estimated from the resistivity peak \cite{Park2011}, the Knight shift anomaly \cite{Lin2015}, and the Hall resistivity \cite{Yang2012} are in good agreement. At ambient pressure and 1 GPa, the onset of the Knight shift anomaly and the Hall resistivity enable one to determine $T^*$, but the peak in the resistivity does not provide a useful estimate of the onset of heavy electron KL scaling behavior at these or other pressures below 1.5 GPa \cite{Yang2008nature}. As we see below, the anomalous behavior of resistivity below 1.5 GPa originates in local moment fluctuations that, however, do not affect the Hall resistivity and the Knight shift. The increase of $T^*$ with pressure seen here appears to be a general characteristic of the Ce-based heavy electron materials.}

\item{The upper boundary of the magnetic quantum critical regime, $T_{QC}$, is determined from the onset of power law scaling of the resistivity \cite{Park2011} and, when the pressure exceeds 1.5 GPa, agrees with the temperature that marks the end of the Kondo liquid scaling at high temperatures \cite{Lin2015}. Curiously, it displays a pressure dependence that is quite similar to that of $T^*$; it is roughly given by $T^*/2$ (as discussed below, the range between $T^*$ and $T_{QC}$ is much larger in CeCoIn$_5$ allowing us to more precisely identify in that case the KL scaling behavior). Below 1.5 GPa, the power law scaling of the resistivity is seen to begin at temperatures large compared to the end of heavy electron scaling behavior at $T^*$; indeed at pressures less than 0.3 GPa, it persists into the local moment regime. This finding, together with the quite similar anomalous behavior of the maximium in the resistivity, tells us that these have a common physical origin, and that both likely reflect local moment fluctuations brought about by the AF QCP.}

\item{A third temperature scale, $T_0$, marks the end of AF quantum critical behavior at pressures less than $\sim2\ $GPa; it is seen in the resistivity \cite{Park2011}, in a pseudo-gap like feature in the spin-lattice relaxation rate, and in peaks in the Hall resistivity measurements \cite{Yang2012}. These behaviors have a common physical origin in the AF QCP, whose fluctuations are seen in the resistivity measurements, while the "relocalization" of heavy electrons it brings about is clearly visible in the Knight shift anomaly \cite{Shirer2012}. At ambient pressure relocalization begins at $T_0\sim2T_N$, which decreases with increasing pressure, and the anomalous behavior of the Hall resistivity follows a very similar pattern. $T_0$ is also seen in the neutron scattering experiments as the onset of precursor magnetic fluctuations of the long-range AF phase and in the  transport measurements as the temperature below which the thermal resistivity and the electrical resistivity deviate from each other \cite{Yang2012}.}

\item{Another crossover temperature scale, $T_X$, marks the lower boundary of the power law scaling in the resistivity at pressures greater than $\sim2.7\ $GPa \cite{Park2011}. It increases with increasing pressure and extrapolates to zero at the QCP. The $T_0$ and $T_X$ lines are candidates for a quantum critical cone that describes the limits of the quantum critical region originating in the magnetic/hybridization QCP. While $T_X$ could mark the delocalization temperature, $T_L$, below which there are no f electron local moments \cite{Yang2012}, since the two types of quantum critical fluctuations are coupled, $T_X$ is likely larger than, but proportional to $T_L$.}

\item{At high pressures, one finds that as one lowers the temperature below $T_X$ the resistivity of the heavy electron liquid first exhibits anomalous behavior brought about by scattering against quantum critical fluctuations; however below a crossover temperature, $T_{FL}$, it exhibits the power law $n=2$ behavior expected for a Fermi liquid \cite{Park2011}. Extrapolations of the $T_{FL}$ line to lower pressures suggests it approaches zero at the QCP, yielding a narrow bandwidth and a heavy effective mass for the heavy electrons \cite{Shishido2005}.}
\end{itemize}

The phase diagram of CeRhIn$_5$ provides a clear illustration of the interplay between the magnetic and hybridization quantum critical fluctuations. What is not shown there is that below 1 K, a slightly different power law scaling is observed at the quantum critical point if one applies a magnetic field large enough to kill the superconductivity \cite{Park2010}. This crossover for the critical exponent may be another indication of the coupling between the magnetic and hybridization QCPs, and is possibly described by the Abrahams-W\"olfle model \cite{Abrahams2012}.

Some desirable further experimental investigations of CeRhIn$_5$ include what changes when the two quantum critical points become separated by doping or other means and the measurement of quantum critical scaling in the Knight shift anomaly at low temperatures.
 
\section{CeCoIn$_5$}
\begin{figure}[t]
\centerline{{\includegraphics[width=.45\textwidth]{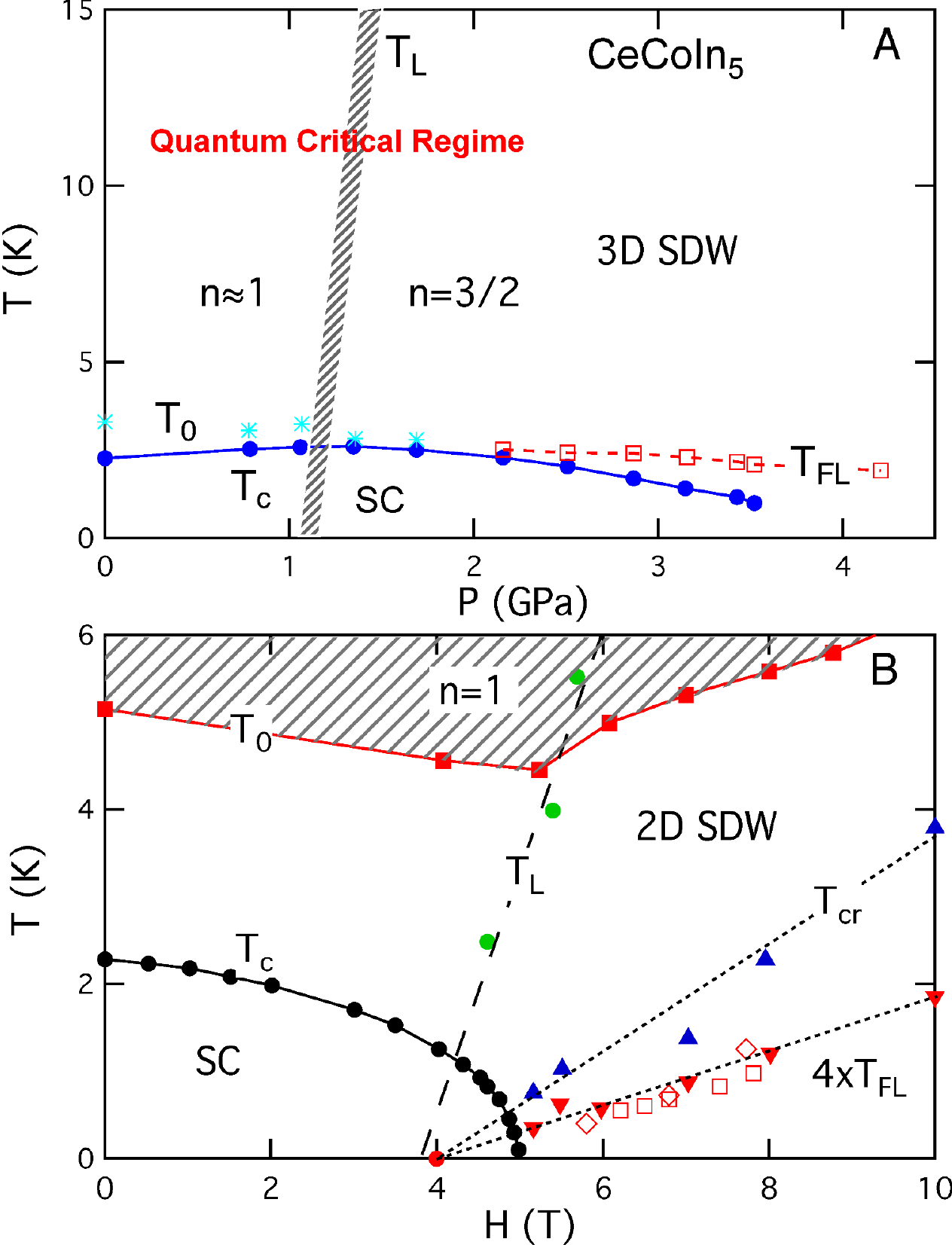}}}
\caption{
{The (A) experimental temperature-pressure and (B) zero-pressure temperature-magnetic field phase diagrams of CeCoIn$_5$ \cite{Sidorov2002,Zaum2011,Paglione2006}. In $A$, $T_L$ marks a proposed boundary (shaded area) between $n=1$ and $n=3/2$ regions \cite{Sidorov2002}; $T_0$ marks the end of the above scaling behavior (stars); $T_{FL}$ is the Fermi liquid temperature determined from the resistivity (squares), and $T_c$ is the superconducting transition temperature (solid circles). The critical pressure is $p_{QC}=1.1\,$GPa. The coherence temperature is $T^*\sim 60\,$K at ambient pressure. In $B$, $T_L$ is the magnetoresistance maximum (green solid circles); $T_{cr}$ is the lower boundary of 2D SDW regime (upward triangles) obtained from thermal expansion measurement; $T_{FL}$ is determined from thermal expansion (downward triangles), Hall effect (open squares), and resistivity measurements (open diamonds) \cite{Zaum2011}. In the shaded area, the resistivity shows ($n=1$) power law scaling and $T_0$ marks its lower boundary (solid squares) \cite{Paglione2006}. Not shown here is a tiny region above $T_{FL}$ where $n=3/2$ power law scaling is seen in the resistivity \cite{Paglione2006}. The critical field is $H_{QC}=4.1\pm0.2\,$T. The slight difference in $T_0$ at zero pressure in $A$ and zero field in $B$ might be due to experimental error in different measurements.
}
\label{fig4}}
\end{figure}

CeCoIn$_5$ is another much studied Class I material in which one can follow the interplay between the magnetic and hybridization quantum critical fluctuations as one reaches a QCP by applying pressure or magnetic field. The experimental results for the phase diagrams showing changes in scaling behavior with pressure and applied magnetic field are shown in Fig.~\ref{fig4} for temperatures far below $T^*\sim 60\,$K  \cite{Sidorov2002,Zaum2011,Paglione2006}:
\begin{itemize}[leftmargin=*]
\item{At pressures around 1.6 GPa, the critical exponent seen in resistivity measurements near to or below $T_{QC}$ changes from $n=1$ to $n=3/2$ \cite{Sidorov2002}. Since the latter corresponds to that expected for a 3D SDW QCP (with "disorder"), we see that high pressure appears to tune the system from 2D to 3D. This crossover of the scaling exponent may correspond to the anticipated delocalization line $T_L$, since 1.6 GPa is not far from the QC pressure of 1.1 GPa \cite{Yang2014}.}

\item{The low-temperature cutoff of the Kondo liquid scaling at ambient pressure seems to be consistent with the onset of resistivity scaling at about $10-20\ $K ($T_{QC}$, not shown in Fig.~\ref{fig4}) \cite{Yang2008,Sidorov2002}. However, considering possible changes in the QCP in the field/pressure phase diagram \cite{Zaum2011}, a comparison between the two may not be proper. In order to establish the interplay between two types of quantum criticality, it will be important to have measurements of Kondo liquid scaling (from the NMR Knight shift) and power law scaling (from the magnetic resistivity) over the entire temperature-pressure/magnetic field phase diagram, following the example of measurements on CeRhIn$_5$.}

\item{In Fig.~\ref{fig4}B, $n=1$ scaling behavior in the resistivity is seen on both sides of the $T_L$ line, but its lower boundary, $T_0$, shows a non-monotonic field-dependence \cite{Paglione2006}. Despite the fact that both sides have the same scaling exponent, it may be argued that the right hand side is governed by itinerant 2D SDW criticality, with an onset temperature that is different for the resistivity and thermal expansion, while the left hand side is governed by the local moment quantum criticality -- possibly an unconventional quantum critical scaling owing to the interplay between magnetic and hybridization quantum critical fluctuations.}

\item{This change of character in the quantum critical scaling takes place at the $T_L$ line which is seen to pass through the point at which $T_0$ changes slope. It plausibly reflects a change of character of the magnetic quantum criticality due to complete hybridization, as shown in the tentative phase diagram for Class I materials in Fig.~\ref{fig2}.}

\item{At high temperatures, the interplay between the magnetic and hybridization fluctuations and its variation with field and pressure have not been well studied. It has been shown at ambient pressure that over a broad temperature region the resistivity is dominated by the scattering of light electrons from isolated local moments. As first noted by Nakatsuji {\sl et al.} \cite{Nakatsuji2004}, at ambient pressure this scattering explains why, as the temperature is reduced, the resistivity first increases, reaches a maximum at $T^*$, and then falls off as $1-f(T)$ until one reaches a temperature $\sim 0.2T^*$. Its change in scaling behavior below this temperature reflects the emerging impact of quantum critical fluctuations on local moment behavior. It is quite possible that it is only at pressures greater than the critical pressure of $\sim 1.1\ $GPa that resistivity measurements begin to tell us about heavy electron behavior at very low temperatures.}

\item{The anomalous Knight shift has been measured at high magnetic fields and shows Kondo liquid scaling below $T^*\sim 60\ $K down to $T_{QC} \sim10\ $K (for field along the $c$-axis) \cite{Curro2004,Yang2008}, behavior that we argue arises from the HY QCP. The NMR spin-lattice relaxation rate also exhibits universal scaling and points to an AF QCP at negative pressure under high magnetic field \cite{Yang2009}.}

\item{In the vicinity of the magnetic-field induced QCP, the intrinsic hybridization parameter, $f_0$, has a power law dependence on the magnetic field, thereby establishing the quantum critical nature of the collective hybridization \cite{Yang2014}.}
\end{itemize}

\section{Discussion and conclusion}
Our proposed framework explains the measured scaling behavior of CeMIn$_5$ and provides insight into that seen in a number of other well studied Kondo lattice materials (see supplementary materials). However, further experiments and analysis are required before we are able to establish more generally the materials for which it is applicable and, for those where it does not apply, understand why it does not. Here are a few open questions that could be answered in future experiments:
\begin{itemize}[leftmargin=*]
\item{The delocalization temperature $T_L$ marks the onset of static hybridization (in the mean-field approach). While indirect evidence for its existence has been obtained in a number of ways, its direct determination is crucial for establishing the range of hybridization quantum critical behavior. This could be done by Knight shift experiments that show a return to one component behavior or by direct measurements of a Fermi surface change across the $T_L$ line using either dHvA or ARPES at finite temperature.}

\item{In many materials, HY and AF QCPs appear to be almost the same. Our framework may be best verified by tuning their relative locations. In YbRh$_2$Si$_2$, this has been done by replacing Rh with Co or Ir \cite{Friedemann2009}, and it will be interesting to check if these replacements lead to the expected change in the quantum critical scaling. Since the critical exponent for the resistivity is not universal \cite{Park2010}, tuning the relative positions of the two QCPs may tell us if this nonuniversality is related to the interplay or competition between the two types of quantum critical fluctuations. In addition, such tuning measurements will provide information on the regions of applicability of the microscopic scaling theory of Abrahams and W\"olfle as the two QCPs are separated.}

\item{While the Knight shift and the resistivity probe quantum critical scaling, direct measurements of the associated quantum critical fluctuations might provide further information. Systematic studies using neutron scattering measurements in the momentum/frequency domain or pump probe technique in the time domain are desirable to establish the existence and interplay of both magnetic and hybridization quantum critical fluctuations.}
\end{itemize}

Theoretically, our proposed scenario may be captured qualitatively by an effective field theory of the Kondo-Heisenberg model. The logarithmic divergence of the Kondo liquid scaling below $T^*$ may be ascribed to a marginal Fermi liquid state due to fluctuations of the hybridization field or an emergent gauge field arising from the Kondo-Heisenberg interaction. It might be possible to develop a physical description based on a (quantum) Ginzburg-Landau model in which the order parameter field, $\phi$, the modulus of which corresponds essentially to the hybridization gap, is imagined to fluctuate in space and time, taking on a well-defined value only in the low temperature limit. We suggest that the variance of the order parameter field, suitably coarse-grained, increases gradually with decreasing temperature starting perhaps well above $T^*$ and grows towards saturation at temperatures of the order of or below $T_{L}$. In the range between $T_{L}$ and $T^*$ the variance takes on intermediate values in keeping with the existence of regions in space and time which are strongly hybridized, forming the heavy fermion fluid, together with other regions in space and time which are weakly hybridized, forming the local-moment fluid in the two-fluid model. For a more complete understanding of $T^*$, it would be necessary to include not only fluctuations of the hybridization field, but also of the emergent gauge field as discussed above \cite{Senthil2004,Paul2007}. In this more complete description, $T^*$ would be associated with the combined effects of the Kondo hybridization term and the Heisenberg intersite interaction term in the Kondo-Heisenberg Hamiltonian of Kondo lattice systems. This collective hybridization may be expected to lead to values of $T^*$ that are quite different from the conventional single-ion Kondo temperature, a prediction supported by a number of detailed studies of Kondo lattice materials \cite{Yang2008nature}. Below $T_{QC}$, the coupling to the spin fluctuations can lead to strong-coupling behavior in which the singularity exceeds that of the marginal Fermi liquid starting point and eventually in the vicinity of AF QCP, give rise to critical quasi-particle behavior as proposed by Abrahams and W\"olfle (see supplementary materials) \cite{Abrahams2012}.

\acknow{Y.Y. was supported by the National Natural Science Foundation of China (grant no. 11522435), the State Key Development Program for Basic Research of China (2015CB921303), the Strategic Priority Research Program (B) of the Chinese Academy of Sciences (XDB07020200) and the Youth Innovation Promotion Association CAS. G.L. acknowledges support from the Engineering and Physical Sciences Research Council (EPSRC grant no. EP/K012894/1) and CNPq/Science without Borders Program. This work was performed in part at the Aspen Center for Physics, which is supported by National Science Foundation (NSF) grant no. PHY-1066293, and in part at the Santa Fe Institute. We thank many colleagues at the Aspen Center for Physics and elsewhere for stimulating discussions.}

\showacknow 



\clearpage


\section*{Supplementary material}

# Supporting Information: Quantum critical scaling and fluctuations in Kondo lattice materials

Yi-feng Yang[a,b,c], David Pines[d], and Gilbert Lonzarich[e]

[a]Beijing National Laboratory for Condensed Matter Physics and Institute of Physics, Chinese Academy of Sciences, Beijing 100190, China; [b]Collaborative Innovation Center of Quantum Matter, Beijing 100190, China; [c]School of Physical Sciences, University of Chinese Academy of Sciences, Beijing 100190, China; [d]Santa Fe Institute, Santa Fe, NM 87501, USA; [e]Cavendish Laboratory, Department of Physics, Cambridge University, Cambridge CB3 0HE, UK

## 1. Probes of scaling and fluctuations

We discuss here the experimental probes that have been used to identify the scaling behavior shown in the main text (1).

**The NMR Knight shift.** While there are a number of candidate experimental signatures of $T^*$ (2), and the scaling behavior of the Kondo liquid found below it (3), only one is unambiguous: the Knight shift experiments which show that above $T^*$ the Knight shift follows the uniform spin susceptibility (and measures local moment behavior) and that it ceases to do this at $T^*$ because a second component, the heavy electron liquid, emerges. As Curro et al. showed (4), below $T^*$ one can disentangle the heavy electron component from the local moment contribution to the Knight shift and show that it displays universal scaling behavior as it varies logarithmically with temperature down to a temperature, $T_{QC}$. Of all the materials that exhibit a Knight shift anomaly that enables one to follow heavy electron behavior directly, the variation of $T^*$ with a control parameter (pressure, applied magnetic field, or doping), has been followed in detail in only one material, CeRhIn$_5$, where pressure is the control parameter (5), while experiments on this and other materials show that $T^*$ does not change appreciably on applying a magnetic field (6).

However, there are some materials, such as YbRh$_2$Si$_2$, where a Knight shift determination of $T^*$ has not proved possible because no anomaly has been observed in the $^{29}$Si Knight shift (7). This could be due to an accidental cancellation of the hyperfine couplings to the two fluids. Since in CeCoIn$_5$ such a cancellation was found at one of the $^{115}$In sites for field along the CeIn$_3$ plane (8), but was not present at other sites or field orientations, it would seem useful to try other field orientations and nuclear sites in YbRh$_2$Si$_2$.

The Kondo liquid heavy electron universal scaling can be cut off at low temperatures by different physical phenomena. Thus for lightly hybridizing materials, the proximity of magnetic order gives rise to heavy electron relocalization at temperatures of $\sim 2T_N$ (9), while for strongly hybridizing materials, magnetic quantum critical fluctuations may act to halt the heavy electron scaling behavior. The latter suggests that the two types of quantum critical fluctuations are coupled, and that future experiments might disclose characteristic heavy electron scaling behavior brought about by magnetic quantum criticality.

**The magnetic resistivity.** At temperatures above $T^*$, the magnetic resistivity is a probe of light conduction electron behavior, and in many materials its temperature behavior, a resistivity that increases as the temperature is decreased, is consistent with these being scattered solely by local moment spin fluctuations. As first noted by Nakatsuji et al. (10) it will then exhibit a maximum at $T^*$ (often called the coherence temperature) because below $T^*$ the emergence of a heavy electron component of strength $f(T)$, is accompanied by a decrease in strength, $1 - f(T/T^*)$, of the local moment component. There are, however, many exceptions to this behavior (3): for example, in YbRh$_2$Si$_2$, there is a broad maximum in the resistivity (likely caused by crystal field effects) that makes it difficult to use it to determine $T^*$; in UBe$_{13}$, the resistivity peak is at $\sim 2.5$ K, far below the coherence temperature $T^* \sim 55$ K determined from other probes.

At temperatures well below $T^*$, the resistivity continues to be dominated by the incoherent scattering of the light conduction electrons, and the emergence of power law scaling ($\rho \sim T^n$, $0 < n < 2$) at $T_{QC}$ is brought about by their scattering by quantum critical fluctuations associated with the AF QCP. The variations in the values of the exponent (e.g., $n = 1.6 \pm 0.2$ in CeIn$_3$ (11), $1.2 \pm 0.1$ in CePd$_2$Si$_2$ (12), 1.0 in YbRh$_2$Si$_2$ and CeCu$_{6-x}$Au$_x$ (13)) suggest the magnetic quantum critical fluctuations possess a variety of physical origins (local moment or itinerant electron AF order, coupled AF and hybridization fluctuations, effects of dimensionality of the magnetic and electronic structure, dimensional crossover, intermediate temperature (non asymptotic) behavior, and more).

At still lower temperatures and slightly above the HY QCP, heavy electron scattering begins to play a role and it seems safe to assume that at $T < T^*/30$, say, the resistivity becomes dominated by heavy electron scattering. A new class of quantum critical fluctuations and a new critical exponent may then emerge when the magnetic quantum critical point is close or coincides with the hybridization quantum critical point, as is the case for YbRh$_2$Si$_2$ and CeCu$_{6-x}$Au$_x$, and a microscopic theory for its behavior has been developed by Abrahams and Wölfle (14). It is likely that different scaling laws will be seen if the two quantum critical points are separated.

In CeCoIn$_5$, we have also used the maximum in the magneto-resistivity to determine the cross-over line $T_L$. We ascribe it to the effects of density fluctuations associated with possible Fermi surface changes.

**The Hall resistivity.** The Hall resistivity is typically very complicated. In heavy electron materials, it has been proposed to originate mainly from the incoherent skew scattering of the light conduction electrons by local moments at high temperatures. Below $T^*$, there can be a contribution from the heavy electrons that might be expected to have a complicated temperature dependence. However, in CeMIn$_5$ materials, the incoherent skew scattering appears to be suppressed, with the result that the coherent contribution from the heavy electrons is seen directly and, moreover, exhibits universal scaling behavior, providing an independent experimental verification of the presence of Kondo liquid scaling (3).

This Hall scaling is also cut off at low temperatures. Since the resistivity can be sensitive to the magnetic quantum critical fluctuations, it is possible that future experiments on the Hall resistivity may disclose new scaling behavior at low temperatures.

**The NMR spin-lattice relaxation rate.** Although the NMR spin-lattice relaxation rate, $1/T_1$, contains distinct local moment and heavy electron contributions that are not always straightforward to disentangle, Yang *et al.* (15) have found that for a number of materials (CeCoIn$_5$, CeRhIn$_5$, PuCoGa$_5$, and URu$_2$Si$_2$), both contributions are so weakly temperature dependent over a considerable range of temperatures, that the observed temperature dependence of $1/T_1$ reflects only that associated with $f(T)$, the strength of the emerging heavy electron component. These results provide another independent measure of their hybridization scaling behavior.

In both CeCoIn$_5$ and CeRhIn$_5$ it has proved possible to use the two-fluid model plus reasonable physical assumptions about the intrinsic local moment contribution to isolate the intrinsic heavy electron contribution, and show that its weak temperature dependence is given by $T_{1KL}T \sim (T + T_0)$, where $T_0$ marks the distance from the magnetic quantum critical point and should not be confused with $T_0$ defined in the paper (3, 16). Since this is of the form expected for heavy electron quantum critical magnetic fluctuations, the result confirms their presence and provides valuable information about the location of the AF QCP for the magnetic fields at which the experiments were carried out.

We note that other probes also provide useful information on heavy electron behavior. For example, point contact spectroscopy provides direct evidence for the emergence of Fano interference between light and heavy electrons at $T^*$ (17, 18), while optical experiments probe the hybridization gap directly (19) and can, in principle, provide more detailed information on changes in hybridization around $T_{QC}$.

## 2. Applications to other materials

Here we discuss the ways in which the framework presented here provides further insight into the behavior of a number of other materials that have been intensively discussed in the literature.

**YbRh$_2$Si$_2$.** In Yb-based materials, in contrast to the Ce-based materials, intrinsic hybridization decreases with increasing pressure; the $T_L$ line therefore has a negative slope in the $(p, T)$ plane. When the magnetic order of YbRh$_2$Si$_2$ is suppressed by applying a small magnetic field, a detailed analysis of its low temperature behavior shows an HY QCP that coincides with its AF QCP, establishing it as a Class I material (6, 20).

- Angle-resolved photoemission spectroscopy (ARPES) experiments (21) have established that for this material, $T^* \sim 50$ K, a result that cannot be confirmed by the presence of a Knight shift anomaly, because in all likelihood, like CeCoIn$_5$ in a planar magnetic field (22), there is an accidental cancellation of the hyperfine couplings (7).

- The behavior of YbRh$_2$Si$_2$ between 50 K and 10 K is complex because its single-ion Kondo temperature is $\sim 20$ K (3); while some have argued that this is the coherence temperature, the ARPES experiment cited above shows it is not.

- Below 10 K a two-fluid analysis of field-induced and pressure-induced changes in magnetostriction, resistivity, Hall measurements, Knight shift, spin-lattice relaxation rate, enabled the authors to establish the presence of power law scaling in the intrinsic hybridization strength, $f_0$, that is produced by the AF quantum critical fluctuations (6). Importantly, it is proved possible to show that the field-dependent Kondo liquid spin-lattice relaxation rate displays behavior that is analogous to that seen in the pressure-dependent Kondo liquid rate for CeCoIn$_5$ (22), namely $T_{1KL}T \sim T + T_0(H)$, where $T_0(H)$ measures the distance from the QCP.

- Between 10 K and 0.3 K the resistivity exhibits QC scaling power law behavior with $n = 1$ (20).

- Below 0.3 K its overall scaling behavior is very well described by the theory of Abrahams and Wölfle based on the premise that it arises from an AF QCP (14).

We also note that Friedemann *et al.* has shown that replacing Rh in YbRh$_2$Si$_2$ by Co or Ir changes it into a Class II or Class III material (23).

**YbAlB$_4$.** For this material there is considerable evidence of a QCP at ambient pressure (24); we propose this is an HY QCP that marks the end of local moment behavior. The negative slope of the $T_L$ line in the $(p,T)$ plane then tells us that at finite pressures one will find a mix of local moment and itinerant heavy electrons; it is therefore plausible that the AF order seen at increased pressure represents local moment antiferromagnetism (25).

**URu$_2$Si$_2$.** Both Knight shift and spin-lattice relaxation rate measurements provide unambiguous evidence for HYQC scaling behavior in URu$_2$Si$_2$. Knight shift experiments reveal anomalous behavior that begins at $T^* \sim 55$ K and displays HYQC scaling down to the hidden order temperature, $T_{HO} = 17.5$ K (9). This behavior is confirmed by measurements of the spin-lattice relaxation rate that shows it scales with $f(T)$ over this same region (15). A two fluid analysis of the static susceptibility suggests that URu$_2$Si$_2$ is a strongly hybridizing material with $f_0 \sim 1.6$, in which case $T_L$ is close to $T_{HO}$, and hidden order is a phase of the fully hybridized heavy electron liquid (16). At very high magnetic fields a proposed phase diagram (26) shows hidden order being suppressed at $\sim 35$ T where it is replaced by SDW-type long-range magnetic order (27). The magneto-resistivity is maximum along a line that ends at $\sim 37 \pm 1$ T (28); if we interpret this as $T_L$, as earlier experiments on CeCoIn$_5$ had suggested, then there is an HY QCP at $\sim 37 \pm 1$ T and $f_0$ must decrease with increasing magnetic field.

**CeCu$_2$Si$_2$.** CeCu$_2$Si$_2$ is a Class II material in which intrinsic hybridization is so strong that the localization QCP is widely separated from AF QCP and lies at substantive negative pressure (29), which implies that over much of the phase diagram one is dealing with only heavy electrons and the AF behavior must be itinerant. In a magnetic field strong enough to suppress superconductivity, its behavior is that expected for an itinerant 3D SDW QCP (30). It has a Knight shift anomaly that begins at $T^* \sim 75 \pm 10$ K and the emergent heavy electron liquid displays Kondo liquid scaling between $T^*$ and 20 K (3, 4), where the resistivity shows a peak (31). We attribute the latter to the loss in local moment spin fluctuations expected if $T_L = 20$ K; we note that such large value for $T_L$ is not surprising since the HY QCP lies within the AF phase. The nuclear quadrupole resonance (NQR) spin-lattice relaxation rate is temperature independent between $T^*$ and $\sim 8$ K, a result that requires that both $T_{1KL}$ and $T_{1SL}$ be temperature independent and very nearly the same (15); if $T_L \sim 20$ K, then below 20 K one is seeing only the KL contribution, which continues to display the $T$-independent scaling behavior expected for AF QC spin fluctuations down to 8 K at which point the AF ordered state spin fluctuations seen in neutron scattering change its behavior.

## 3. On the nature of the Kondo liquid state on the border of a magnetic transition

Our aim is to search for a microscopic understanding of the Pines-Yang two-fluid model of a system of conduction electrons interacting with localized f moments as described by the Kondo-Heisenberg Hamiltonian (1, 19).

Here we focus attention on the evolution with temperature of the Kondo liquid state at the quantum critical point separating the magnetic and non-magnetic state as a function of a tuning parameter such as pressure. Below the coherence temperature $T^*$, the coupling of the conduction electrons and f electrons leads to the beginning of on-site hybridization and inter-site spin correlations described, respectively, by the Kondo and Heisenberg terms of the Hamiltonian.

Studies that indicate that $T^*$ is related to the Heisenberg interaction parameter, $J_H$, together with optical conductivity and other measurements of the electronic structure, suggest that both inter-site f electron spin correlations and conduction electron-f electron hybridization develop below $T^*$, at least in systems of primary interest that are on the border of AF QCPs (2).

The degree of hybridization fluctuates in space and time and may be described by a quantum field of the form $\phi = c^+ f$, where $c$ and $f$ correspond, respectively, to the conduction electron and f electron field operators with the same site and spin index. The fluctuations in $\phi$ lead to time-varying regions that may be described as essentially hybridized and hence characterized by heavy quasiparticle bands, and other regions in which hybridization is relatively ineffective and hence characterized by local f electron moments and conduction electron bands. This gives rise to the idea of two coexisting fluids consisting of heavy quasiparticles on the one hand and local moments (together with conduction electrons) on the other.

The Pines-Yang two-fluid model describes the Kondo liquid state below $T^*$ near the quantum critical point in terms of temperature-dependent volume fractions and thermodynamic and transport properties for the two fluids. The local moment fluid is dominant near to $T^*$, while the quasiparticle fluid is dominant in the low temperature limit.

The chief characteristic of the heavy quasiparticle fluid is a marginal Fermi liquid $T\ln(T^*/T)$ heat capacity, $C$, and $T$-linear electrical resistivity, $\rho$, extending down to a cross-over "quantum critical" temperature $T_{QC}$ below $T^*$. Typically well below $T_{QC}$ the departure from Fermi liquid behavior is still more dramatic and in the Abrahams-Wölfle model for YbRh$_2$Si$_2$ (14) the heat capacity and electrical resistivities both vary as $T^{3/4}$, making the ratios $C/T \sim 1/T^{1/4}$ and $(\rho - \rho_0)/T^2 \sim 1/T^{5/4}$ more singular than that of the marginal Fermi liquid state.

This can be understood in a unified way in terms of the effects of quasiparticle-quasiparticle interactions in the hybridized state. In the simplest case these interactions may be represented in terms of the exchange of spontaneous spin fluctuations. Additional analogous contributions can also arise from fluctuations of the hybridization field $\phi$ as well as the "bond" field $\chi = f_i^+ f_j$, where $f_i$ and $f_j$ correspond to f electron field operators at nearest neighbor sites with the same spin index.

The above non-Fermi liquid behavior can be understood intuitively in terms of a phenomenological model analogous to that of electrons interacting with phonons, but with phonons replaced by dissipative spin fluctuations characterized by a wavevector and frequency dependent magnetic susceptibility of the RPA form as employed, for example, in the spin-fermion model.

In the weak coupling limit, this form of the generalized susceptibility leads to a relaxation spectrum of magnetic fluctuations of the form $\Gamma_q \sim q^z$ where the dynamical exponent, $z$, is equal to 2 or 3 for antiferromagnetic and ferromagnetic critical

fluctuations, respectively. This leads directly to the marginal Fermi liquid model behavior on the border of ferromagnetism in 3D and antiferromagnetism in 2D, provided in the latter case that "impurities" lead to a suppression of Fermi surface anisotropy induced by spin fluctuation scattering at the antiferromagnetic ordering wavevector.

The growth in the mass of the quasiparticles with decreasing temperature (as measured by the growth of $C/T$ with decreasing $T$) can, at sufficiently low temperatures, lead to a strong modification of the spin fluctuation spectrum due to feedback from the strong renormalization of the quasiparticle spectrum, an effect omitted in the weak coupling approximation.

In the strong coupling description, the parameters in the wavevector and frequency dependent susceptibility acquire renormalization factors tracking the frequency dependence quasiparticle mass, leading to a modification of the dynamical exponent $z$. A self-consistent solution of the renormalization of both the quasiparticle and the spin fluctuation spectra leads, on the border of antiferromagnetism in 3D, to a non-Fermi liquid state that, as described above in the Abrahams-Wölfle model (14), has a more singular mass renormalization than that of the marginal Fermi liquid state.

The above findings can be interpreted in a minimal way as follows. We begin with the weak-coupling RPA susceptibility of the form

$$\chi_{q\omega} = \frac{\alpha_{q\omega}}{1 - \lambda \alpha_{q\omega}},$$
$$\alpha_{q\omega} = \alpha\left[1 - \frac{(q-Q)^2}{k_0^2} + \frac{i\omega}{\nu_0 q}\right], \quad [1]$$

so

$$\chi_{q\omega} \approx \frac{\alpha}{1 - \lambda\alpha + \frac{\lambda\alpha(q-Q)^2}{k_0^2} - \frac{i\lambda\alpha\omega}{\nu_0 q}}. \quad [2]$$

where $\alpha_{q\omega}$ is the starting non-interacting susceptibility (or Lindhard function) for wavevector $q$ and frequency $\omega$, $\alpha$ is the starting density of states at the Fermi level, $k_0$ and $\nu_0$ are of the order of the Fermi wavevector and starting Fermi velocity, respectively, $Q$ is the ordering wavevector and $\lambda$ is the coupling constant. In this case when $Q$ is non zero the dynamical exponent at the quantum critical point defined by $\lambda\alpha \to 1$ is $z = 2$ and the thermal population of spin fluctuations is expected to be of order $(q_T)^d$, where $d$ is the spatial dimension and $q_T$ is the characteristic thermal wavevector defined by $(q_T)^z \sim T$. This leads to $C/T \sim 1/T^{(z-d)/z}$, which plausibly has a marginal Fermi liquid (logarithmic) form when the exponent $(z-d)/z$ tends to zero. This may be confirmed for the case of a 2D antiferromagnetic quantum critical point in the presence of impurity scattering as stated above. In the case of critical fluctuations where $Q \to 0$, the above form for $\chi_{q\omega}$ yields $z = 3$ and marginal Fermi liquid behavior may be expected to arise in 3D. This is relevant to ferromagnetic as well as $\phi$ or $\chi$ field critical fluctuations that may also be important in the Kondo liquid state.

In a strong-coupling description we expect $\nu_0$ and $1/\alpha$ to be scaled by the inverse mass renormalization factor, $Z$, assumed to be isotropic. This does not apply, however, to the $(q-Q)^2$ term which arises from contributions away from the Fermi energy where renormalization is relatively weak. Thus at the quantum critical point where $\lambda\alpha/Z \to 1$ the renormalized susceptibility becomes

$$\chi_{q\omega} \approx \frac{\alpha}{\frac{Z^2(q-Q)^2}{k_0^2} - \frac{i\omega}{\nu_0 q}}. \quad [3]$$

If $Z \sim \omega^\theta$ then we see that the renormalized dynamical exponent is $z_{ren} = z/(1-2\theta)$, where $z = 2$ in the above case. Assuming that the argument given above for the thermal population continues to apply, then $C/T \sim 1/T^{(1-d/z_{ren})}$. Matching for self-consistency this expression for $C/T$ to the mass renormalization factor $1/Z \sim 1/T^\theta$ for the case $d = 3$ & $z = 2$ yields $\theta = 1/4$, so that $C/T \sim 1/T^{1/4}$. Thus the quasiparticle mass divergence with decreasing $T$, as measured by $C/T \sim 1/Z$, is expected to be more singular than that of the marginal Fermi liquid as stated above.

A remarkable finding that is important for the self-consistency of the Abrahams-Wölfle model, is that although the thermodynamic and transport properties depart strongly from those of a Fermi liquid and the overlap between a quasiparticle state and a single electron state vanishes ($Z \to 0$ on the Fermi surface), a description in terms of well-defined heavy quasiparticles survives even in the strong coupling regime. Also, importantly, the Fermi surface associated with these exotic quasiparticles is that corresponding to the hybridized bands and hence encloses a volume satisfying Luttinger's theorem including both conduction and f electron contributions.

A unified description that may be appropriate for nearly antiferromagnetic metals such as the tetragonal YbRh$_2$Si$_2$ and the Ce115 compounds that have strongly anisotropic, quasi-2D, spin fluctuation spectra may be as follows. At elevated temperatures where $T$ exceeds the bandwidth of spin fluctuations along the $c$-axis, a quasi-two dimensional description may be appropriate. This (along with any 3D ferromagnetic, $\phi$ or $\chi$ field critical fluctuations present) leads to a marginal Fermi liquid description as indicated above. At sufficiently low temperatures where $T$ falls below the bandwidth of spin fluctuations along the $c$-axis, the spin fluctuations are effectively described by a 3D model and, provided that the mass renormalization becomes sufficiently high to lead to strong coupling behavior ($T \ll T_{QC}$), the thermal and transport properties becomes more singular than expected for the marginal Fermi liquid state. In particular the quasiparticle mass is characterized by a $1/T^{1/4}$ rather than a marginal Fermi liquid $\ln(T^*/T)$ divergence in the low $T$ limit. This microscopic description of the Kondo liquid state based on the Abrahams-Wölfle model (14) is broadly consistent with the Pines-Yang two-fluid phenomenology (16) and appears to be at least qualitatively consistent with observations in the above classes of Kondo systems.



\end{document}